\documentclass[12pt]{article}
\usepackage{latexsym}
\def\Dslash{D \hspace{-.66em} / \hspace{.14em}}
\def\dslash{\partial \hspace{-.54em} / \hspace{.09em}}
\def\mbf#1{\mbox{\boldmath $#1$}}
\makeatletter
        
        \@addtoreset{equation}{section}
\makeatother
\begin{document}
\rightline{YAMAGATA-HEP-99-22}
\rightline{Nov 1999}
\vspace{1.0truecm}
        \renewcommand{\thefootnote}{\fnsymbol{footnote}}
\centerline{\Large\bf 
Gaugeon Formalism for Spin-3/2}
\centerline{\Large\bf 
 Rarita-Schwinger Gauge Field
}
\vspace{1truecm}
\centerline{Ryusuke ENDO and Minoru Koseki$^{*}$}
\bigskip
\centerline{\it Department of Physics, 
Yamagata University, Yamagata 990-8560, Japan}
\centerline{\it $^{*}$Graduate School of Science and Technology, 
Niigata University } 
\centerline{\it Niigata 950-2181, Japan}
\vspace{1truecm}
\centerline{\bf Abstract}
\bigskip
We provide a gauge covariant formalism of the canonically quantized 
theory of spin-3/2 Rarita-Schwinger gauge field. 
The theory admits a quantum gauge transformation by which we can 
shift the gauge fixing parameter.
The quantum gauge transformation does not change the BRST charge. 
Thus, the physical Hilbert space is trivially independent of 
the gauge fixing parameter. 


\section{Introduction}
In the standard formalism of canonically quantized gauge theories\cite{N,KO}
we do not consider the gauge transformation freely.
There are no quantum gauge freedom since the quantum 
theory is defined only after the gauge fixing. 

Yokoyama's gaugeon formalism\cite{OEM,YK,txt,YM1,Smatrix,YM2} 
provides a wider 
framework in which we can consider the quantum gauge transformation 
among a family of Lorentz covariant linear gauges. In this formalism 
a set of extra 
fields, so called gaugeon fields, 
is introduced as the quantum gauge freedom. 
This theory was proposed for the quantum 
electrodynamics\cite{OEM,YK,txt} 
and for the Yang-Mills theory.\cite{YM1,YM2} \  
Owing to the quantum gauge freedom it becomes almost trivial 
to check 
the gauge parameter independence of the physical $S$-matrix.%
\cite{Smatrix}

We should ensure that the gaugeon modes do not contribute to 
the physical processes. In fact, the gaugeon fields yield negative 
normed states.\cite{OEM} \  
To remove these unphysical 
modes Yokoyama 
imposed a Gupta-Bleuler type subsidiary condition,\cite{OEM,YM1,YM2} \ 
which is not applicable if interaction exists for gaugeon fields. 
Yokoyama's 
condition has been improved by introducing
BRST symmetry for gaugeon fields.\cite{Abe,Izawa,KSE,KSE_YM} \ 
Unphysical gaugeon modes, as well as unphysical modes of the 
gauge field, are removed by the single Kugo-Ojima type condition.\cite{KO} \ 
Thus, the formalism is now applicable even in the 
background gravitational field. 
The BRST symmetry is also very helpful in the
analysis of the gauge structure of the Fock space in  
the gaugeon formalism.\cite{KSE,RE}  \ 

By now, we have the BRST symmetric gaugeon formalism of 
the electromagnetic gauge theory\cite{Izawa,KSE} and of the 
Yang-Mills gauge theory.\cite{Abe,KSE_YM} \ 
There are, however, other types of gauge fields, 
such as the gravitational field, the gravitino (spin-3/2 gauge field), 
the anti-symmetric tensor gauge fields and the string theory.
One might wonder whether the gaugeon formalism is applicable 
to these gauge fields. 
In the present paper, 
we formulate a BRST symmetric gaugeon formalism 
for the spin-3/2 Rarita-Schwinger gauge field. Although we treat 
it mainly in the free field case, we can straightforwardly 
incorporate the interaction with the Ricci flat background gravitational 
field. 

The paper is organized as the following. 
In \S2, we briefly review the theory of Hata and Kugo\cite{HK} \ 
as a standard formalism of the canonically quantized spin-3/2 
gauge field. In \S3, we propose a BRST symmetric gaugeon formalism 
for the spin-2/3 gauge field, where gauge fixing parameter 
can be shifted by a $q$-number gauge transformation. 
We see in \S4 how the Fock space of the standard formalism 
is embedded in the wider Fock space of the present formalism. 
Section 5 is devoted to comments and discussion, including the remarks 
on other types of gaugeon formalism.

\section{Standard formalism}
The classical Lagrangian of the free gravitino field $\psi_\mu$
in $n$ ($\geq 3$) dimensional flat space-time is given by%
           \footnote{%
                      We use the convention of Bjorken-Drell. For 
                      example, 
                      $\{\gamma_\mu,\,\gamma_\nu \} = 2g_{\mu\nu}$
                      with 
                      $g_{\mu\nu}=\mbox{diag}(1,-1,-1,\dots,-1)$.
                   } 
\begin{equation}
{\cal L}_{\mbox{\scriptsize RS}}
                =
           -{i\over 2}\, \bar \psi_\mu \gamma^{\mu \nu \lambda}
              \partial _\nu \psi_\lambda ,
\label{Eq:L_RS}
\end{equation}
where $\gamma^{\mu\nu\lambda}$ is the matrix 
$\gamma^\mu \gamma^\nu \gamma^\lambda$ 
anitisymmetrized with respect to 
$\mu$, $\nu$ and $\lambda$, 
\begin{equation}
   \gamma^{\mu\nu\lambda}
   = {1\over 6}( \gamma^\mu \gamma^\nu \gamma^\lambda \pm \mbox{5 terms}).
\end{equation}
The factor 1/2 arises in (\ref{Eq:L_RS}) since we assumed 
the field $\psi_\mu$ to be a Majorana spinor-vector.%
          \footnote{%
                    If one needs to consider the non-Majorana case, 
                    the factor 1/2 should be omitted.  
                    In the following in this paper, we assume that
                    all the spinor fields are Majorana. 
                   } 
The Lagrangian (\ref{Eq:L_RS}) is invariant 
up to total derivatives under the gauge transformation
\begin{equation}
\delta \psi_\mu = \partial_\mu \Lambda, 
\end{equation}
where $\Lambda$ is an arbitrary 
spinor field.

To carry out the quantization, it is necessary to add a gauge fixing 
term and a corresponding Faddeev-Popov (FP) ghost term. 
As a standard formalism, we use the theory  
of Hata and Kugo.\cite{HK} \ 
Their quantum Lagrangian is given by 
\begin{equation}
{\cal L}_{\mbox{\scriptsize HK}}
              =
              {\cal L}_{\mbox{\scriptsize RS}}
              +
              \bar B \dslash (\gamma \psi) 
              -
               {ia \over 2} \bar B \dslash B
              -i 
              \partial_\mu \bar c_*  \partial^\mu c,
\label{Eq:L_HK}              
\end{equation}
where  $\gamma \psi=\gamma^\mu \psi_\mu$, 
$\dslash=\gamma^\mu \partial_\mu$, 
$B$ is a 
spinor multiplier 
(subject to the Fermi statistics), 
$c$ and $c_*$ are the 
spinor 
FP ghosts (subject to the Bose statistics), 
and  
$a$ is a numerical gauge fixing 
parameter.\footnote{%
                   Hereafter, we refer to this parameter $a$ as the
                   standard gauge parameter.
                   } 
Note that the FP ghost fields satisfy the second order differential 
equation. Owing to this property, FP ghosts $c$ and $c_*$ together with 
the multiplier $B$ realize the correct 
ghost counting.\cite{NK,EK} \ 

The field equations are given by
  \begin{eqnarray}
   && -i \gamma^{\mu\nu\lambda} \partial_\nu \psi_\lambda
       = \gamma^\mu \dslash B ,
                          \nonumber \\
   && \dslash (\gamma \psi)
      = i a  \dslash  B,
                          \nonumber
                          \\
   && \Box c = \Box c_* = 0,
   \label{Eq2:standard}
  \end{eqnarray}
from which we also have
  \begin{eqnarray}
  &&  \Box B=0,
                          \nonumber 
                          \\
  && \Box (\gamma \psi)=0.
                          \label{Eq2:Bgammapsi}
  \end{eqnarray}

The Lagrangian (\ref{Eq:L_HK}) leads to the following
$n$-dimensional (anti)commutation relations: 
\begin{eqnarray}
&& \{ \psi_\mu(x), \bar \psi_\nu(y) \}
       = \left[ 
                g_{\mu\nu} \dslash 
                 + {1 \over n-2} 
                            \gamma_\mu \dslash \gamma_\nu
                 - {2\over n-2} 
                            \left(
                                  \gamma_\mu \partial_\nu
                                + \gamma_\nu \partial_\mu
                           \right)
         \right] D(x-y) 
                                      \nonumber \\
     && 
         \hspace{7em} 
        + \left(  {4\over n-2}-a  \right)                    
             \partial_\mu \partial_\nu \dslash E(x-y) ,
  \nonumber 
  \\
&&  \{ B(x), \bar\psi_\nu(y)\}
                = i\partial_\nu D(x-y),
  \nonumber 
  \\
&& \{ B(x), \bar B(y) \}  =  0,
  \nonumber 
  \\
&& \left[\, c(x), \bar c_*(y) \,\right]
                     = - D(x-y),
  %
\label{CRs2}
\end{eqnarray}
where the functions $D$ and $E$ are defined by
\begin{eqnarray}
%
%
&&   \Box D(x)=0, \qquad D(0,{\mbf x})=0, 
                \qquad \dot D(0,{\mbf x})= - \delta^{(n-1)}({\mbf x}),
\nonumber 
\\
%
%
&&   \Box E(x) = D(x), \qquad E(0,{\mbf x}) = \dot E(0,{\mbf x}) =0. 
\label{Eq2:DandE}
\end{eqnarray}
From the first equation of (\ref{CRs2}) we have
\begin{equation}
 \{ \gamma \psi(x), \bar \psi(y) \gamma \}
                = -a \dslash D(x-y),
  \label{CR2:gammapsi}
\end{equation}
There are two special gauges. 
One is the Landau gauge ($a=0$), in which 
$\{\gamma \psi, \bar \psi \gamma \}=0$ so that 
the $\gamma \psi$ mode has vanishing norm. 
The other is the Feynman gauge ($a=4/(n-2)$), in which
$\psi_\mu$ does not include dipole modes: the Feynman propagator 
$\langle T(\psi_\mu \bar \psi_\nu) \rangle$ 
does not have $1/p^4$ term.%
    \footnote{%
              In the Feynman gauge, it is convenient to use  a
              field variable 
              $
                 \phi_\mu = \psi_\mu
                           - {1\over 2} \gamma_\mu (\gamma \psi). 
              $%
              \cite{ET} \  
              When $a=4/(n-2)$, 
              it satisfies the Dirac equation  
              $\dslash \phi_\mu =0$,  
              and the anticommutation  relation becomes 
              $$
               \{\phi_\mu(x), \bar \phi_\nu(y) \}
                 = g_{\mu\nu} \dslash D(x-y).
              $$
            }

The Lagrangian (\ref{Eq:L_HK}) is invariant up to total 
derivative terms under the following BRST transformation:
\begin{eqnarray}
&&   \delta_{\mbox{\scriptsize B}} \psi_\mu
                 = i \partial_\mu c, 
                 \nonumber 
                 \\
&&   \delta_{\mbox{\scriptsize B}} c_* = B, 
                 \nonumber
                 \\
&&   \delta_{\mbox{\scriptsize B}} B
     = \delta_{\mbox{\scriptsize B}} c =0.
\label{Eq2:BRST}
\end{eqnarray}
The corresponding conserved BRST charge is given by
\begin{equation}
    Q_{\mbox{\scriptsize B(HK)}}
    =
    -i \int  \bar B 
             \stackrel{\leftrightarrow}{\partial}_0
              c \,\,d^{n-1} {\mbf x},  
\label{Eq2:QB}
\end{equation}
where 
$
            \stackrel{\leftrightarrow}{\partial}_0
              = \partial_0
                - \stackrel{\leftarrow}{\partial}_0
$.
Using this charge we can define the physical subspace 
${\cal V}_{\mbox{\scriptsize phys}}^{\mbox{\scriptsize (HK)}}$ 
as a space of the states which satisfy 
the physical subsidiary condition of Kugo-Ojima,\cite{KO}
\begin{equation}
      Q_{\mbox{\scriptsize B(HK)}} 
      \left|\mbox{phys} \right\rangle =0.
\end{equation}

There are many unphysical zero-normed states 
in the physical subspace 
${\cal V}_{\mbox{\scriptsize phys}}^{\mbox{\scriptsize (HK)}}$.
In fact, 
${\cal V}_{\mbox{\scriptsize phys}}^{\mbox{\scriptsize (HK)}}$ 
has a zero-normed subspace 
$$
\mbox{Im} \, Q_{\mbox{\scriptsize B(HK)}}
 =
 \left\{  |\Phi\rangle ; \,
          |\Phi\rangle = 
          Q_{\mbox{\scriptsize B(HK)}}|*\rangle
 \right\}.
$$
Considering the quotient space of 
${\cal V}_{\mbox{\scriptsize phys}}^{\mbox{\scriptsize (HK)}}$ 
by 
this 
subspace, we
can define the  physical Hilbert space, 
\begin{equation}
  {\cal H}_{\mbox{\scriptsize phys}}^{\mbox{\scriptsize (HK)}}
  =
  {\cal V}_{\mbox{\scriptsize phys}}^{\mbox{\scriptsize (HK)}}
   /
   \mbox{Im}\, Q_{\mbox{\scriptsize B(HK)}}, 
\end{equation}
which has positive definite metric.

\section{Gaugeon formalism}
We start from the Lagrangian
\begin{eqnarray}
{\cal L} &=& {\cal L}_{\mbox{\scriptsize RS}}
          + \bar B \dslash (\gamma \psi) 
          - {i\varepsilon \over 2}
                ( \bar Y_* + \alpha \bar B)
                     \dslash (Y_* + \alpha B)     
          - \partial_\mu \bar Y_* \partial^\mu Y
          \nonumber \\ &&
          -i \partial_\mu \bar c_*  \partial^\mu c
          -i \partial_\mu \bar K_*  \partial^\mu K,
\label{Eq:L_ggon}
\end{eqnarray}
where, in addition to the usual multiplier $B$ and FP ghosts
$c$ and $c_*$, 
we have introduced 
spinor gaugeon fields 
$Y$ and $Y_*$ (subject to the Fermi statistics) and 
corresponding spinor FP ghosts 
$K$ and $K_*$ (subject to the Bose statistics). 
In (\ref{Eq:L_ggon}), $\varepsilon$ denotes a sign factor 
($\varepsilon = \pm 1$) and $\alpha$ is a numerical gauge 
fixing parameter. 
As seen below, the standard gauge fixing parameter, 
which is denoted by $a$ in the present paper, 
can be identified with
\begin{equation}
       a= \varepsilon \alpha^2.
              \label{Eq3:gaugeparameter}
\end{equation} 

\subsection{Field equations and (anti)commutation relations}
Field equations which follow from (\ref{Eq:L_ggon}) are
\begin{eqnarray}
&& -i \gamma^{\mu\nu\lambda} \partial_\nu \psi_\lambda
    = \gamma^\mu \dslash B ,
             \nonumber 
             \\
&& \dslash (\gamma \psi)
   = i\varepsilon \alpha \dslash (Y_* + \alpha B),
                         \nonumber 
                         \\
&&  \Box Y 
    = i \varepsilon \dslash (Y_* + \alpha B),
                         \nonumber 
                         \\
&& \Box Y_* =0
                         \nonumber %
                         \\
&& \Box c = \Box c_* = 0, 
                         \nonumber \\
&& \Box K = \Box K_*=0. 
\label{Eq3:fieldequation}
\end{eqnarray}
From these equations we also have
\begin{eqnarray}
&&  \Box B=0,
                \nonumber 
                \\
&&  \Box (\gamma \psi)=0,     
                \nonumber 
                \\
&&  \dslash \Box Y=0.
                \label{Eq3:Y}
\label{Eq3:fieldequation2}
\end{eqnarray}

The canonical prescription of quantization leads to
the following $n$-dimensional (anti)commutation relations:
Among the usual fields 
($\psi_\mu$, $B$, $c$, $c_*$), 
we have
\begin{eqnarray}
&& \{ \psi_\mu(x), \bar \psi_\nu(y) \}
       = \left[ 
                g_{\mu\nu} \dslash 
                 + {1 \over n-2} 
                            \gamma_\mu \dslash \gamma_\nu
                 - {2\over n-2} 
                            \left(
                                  \gamma_\mu \partial_\nu
                                + \gamma_\nu \partial_\mu
                           \right)
         \right] D(x-y) 
                      \nonumber \\
     && 
         \hspace{7em} 
        + \left(  {4\over n-2}- \varepsilon \alpha^2  \right)                   
             \partial_\mu \partial_\nu \dslash E(x-y) ,
  \nonumber 
  \\
&&  \{ B(x), \bar\psi_\nu(y)\}
                = i\partial_\nu D(x-y),
  \nonumber
  \\
&& \{ B(x), \bar B(y) \}  =  0,
  \nonumber 
  \\
&& [ \, c(x), \bar c_*(y) \,] = - D(x-y).
  \label {CR3:standard} 
\end{eqnarray}
Among the gaugeons and 
their FP ghosts ($Y$, $Y_*$, $K$,  $K_*$), 
we have
\begin{eqnarray}
&& \{ Y_*(x), \bar Y_*(y) \}  =  0,
  \nonumber 
  \\
&& \{ Y_*(x), \bar Y(y) \}  
               =  -i D(x-y),
  \nonumber
  \\
&& \{ Y(x), \bar Y(y) \}  
              =  \varepsilon \dslash E(x-y),
  \nonumber
  \\
&& [\, K(x), \bar K_*(y) \,] = -D(x-y).              
  \label{CR3:gaugeons}
\end{eqnarray}
Anticommutators between the gaugeons and the usual fields
are given by
\begin{eqnarray}
&& \{ Y_*(x), \bar B(y) \} = \{ Y_*(x), \bar \psi_\mu(y) \}  =  0,
  \nonumber
  \\
&& \{ Y(x), \bar B(y) \}  =  0,
  \nonumber
  \\
&& \{ Y(x), \bar \psi_\mu(y) \} 
               =  - \varepsilon \alpha \partial_\mu 
                                        \dslash E(x-y).
  \label{CR3:gaugeon-standard} 
\end{eqnarray}
The (anti)commutation relations (\ref{CR3:standard}) are exactly 
the same with (\ref{CRs2})  
if we assume (\ref{Eq3:gaugeparameter}). 
In particular, 
$\alpha=0$ corresponds to the Landau gauge, and
$\alpha=2/\sqrt{n-2}$ (together with $\varepsilon=+1$) leads to
the Feynman gauge. 
We notice from (\ref{CR3:gaugeon-standard}) that
in the Landau gauge ($\alpha = 0$) 
the gaugeon modes $Y$ and $Y_*$ 
completely decouple from the usual fields 
$\psi_\mu$ and $B$.

%
\subsection{BRST symmetry}
The Lagrangian (\ref{Eq:L_ggon}) is invariant 
up to total derivatives 
under the following BRST transformation:
\begin{eqnarray}
&&   \delta_{\mbox{\scriptsize B}} \psi_\mu
                 = i \partial_\mu c, 
                 \nonumber 
                 \\
&&   \delta_{\mbox{\scriptsize B}} c_* = B, 
                 \nonumber
                 \\
&&   \delta_{\mbox{\scriptsize B}} B
     = \delta_{\mbox{\scriptsize B}} c =0,
                 \nonumber 
                 \\
&&   \delta_{\mbox{\scriptsize B}} Y = iK, 
                 \nonumber
                 \\
&&   \delta_{\mbox{\scriptsize B}} K_* = Y_*, 
                 \nonumber
                 \\
&&   \delta_{\mbox{\scriptsize B}} Y_* 
      =  
      \delta_{\mbox{\scriptsize B}} K = 0, 
\label{Eq3:BRST}
\end{eqnarray}
which obviously satisfies the nilpotency, 
${\delta_{\mbox{\scriptsize B}}}^2=0$. 
%
The corresponding conserved BRST charge is given by
\begin{equation}
    Q_{\mbox{\scriptsize B}}
    =
    -i \int  \left(
                   \bar B \stackrel{\leftrightarrow}{\partial}_0 c 
                   +
                   \bar Y_* \stackrel{\leftrightarrow}{\partial}_0 K
             \right)
                    \,d^{n-1} {\mbf x}. 
\label{Eq3:Q_B}
\end{equation}
By the help of this charge we can define the physical subspace 
${\cal V}_{\mbox{\scriptsize phys}}$ as a space of the states 
satisfying 
\begin{equation}
      Q_{\mbox{\scriptsize B}} 
      \left|\mbox{phys} \right\rangle =0.
\label{Eq3:subsidiary}      
\end{equation}
This subsidiary condition removes the gaugeon modes as well 
as the unphysical gravitino modes from the physical subspace; 
$Y$ and $Y_*$ together with $K$ and $K_*$ constitute 
the BRST quartet.\cite{KO} 

\subsection{$q$-number gauge transformation}
The Lagrangian (\ref{Eq:L_ggon}) admits a $q$-number 
gauge transformation. Under the field redefinition 
\begin{eqnarray}
&&   \hat \psi_\mu = \psi_\mu + \tau \partial_\mu Y,
                \nonumber \\
&&   \hat Y_* = Y_* - \tau B,
                \nonumber \\
&&   \hat B = B, \qquad   \hat Y = Y,
                \nonumber \\
&&   \hat c = c + \tau K, 
                \nonumber \\
&&   \hat K_* = K_* - \tau c_*,
                \nonumber \\
&&   \hat c_* = c_*, \qquad   \hat K = K,
\label{Eq3:qnumbergauge}
\end{eqnarray}
with $\tau$ being a numerical parameter, 
the Lagrangian (\ref{Eq:L_ggon}) 
is {\it form invariant}\/ (up to total derivative terms), 
that is, it satisfies
\begin{equation}
      {\cal L}(\varphi_A, \alpha)
      =
      {\cal L}(\hat \varphi_A, \hat \alpha)
      + \mbox{total derivatives},
\label{Eq3:forminvariance}
\end{equation}
where $\varphi_A$ stands for any of the relevant fields 
and $\hat \alpha$ is defined by
\begin{equation}
  \hat \alpha = \alpha + \tau.
\label{Eq3:alphahat}
\end{equation}

An immediate conclusion from 
the form invariance (\ref{Eq3:forminvariance}) 
is the following: 
All the field equations 
and all the (anti)commutation relations
are {\it gauge covariant}\/ 
under the $q$-number gauge transformation (\ref{Eq3:qnumbergauge}): 
$\hat \varphi_A$ satisfies the field equations 
(\ref{Eq3:fieldequation}), 
(\ref{Eq3:fieldequation2}) and the (anti)commutation relations
(\ref{CR3:standard}) $\sim$ (\ref{CR3:gaugeon-standard}) 
with $\alpha$ replaced by $\hat \alpha$.

It should be noted that the $q$-number gauge transformation 
(\ref{Eq3:qnumbergauge}) commutes with the BRST transformation 
(\ref{Eq3:BRST}). As a result, our BRST charge 
(\ref{Eq3:Q_B}) is invariant under the $q$-number gauge 
transformation:
\begin{equation}
   \widehat Q_{\mbox{\scriptsize B}}
   =
   Q_{\mbox{\scriptsize B}}.
\end{equation}
The physical subspace 
${\cal V}_{\mbox{\scriptsize phys}}$ 
is, therefore,
invariant under the $q$-number gauge transformation:
\begin{equation}
  \widehat  {\cal V}_{\mbox{\scriptsize phys}}
        = {\cal V}_{\mbox{\scriptsize phys}}.
\end{equation}
Similarly, our physical Hilbert space 
$
    {\cal H}_{\mbox{\scriptsize phys}}
     =
       {\cal V}_{\mbox{\scriptsize phys}}
       /
       \mbox{Im}\, 
       Q_{\mbox{\scriptsize B}}
$
is also gauge invariant: 
\begin{equation}
  \widehat  {\cal H}_{\mbox{\scriptsize phys}}
        = {\cal H}_{\mbox{\scriptsize phys}}.
\end{equation}

%

\section{Gauge structure of the Fock space}
As well as the BRST symmetry (\ref{Eq3:BRST}), the 
Lagrangian (\ref{Eq:L_ggon}) has several other symmetries. 
In particular, we have the following BRST-like conserved charges: 
\begin{eqnarray}
&& Q_{\mbox{\scriptsize B(HK)}}
         = -i \int \bar c 
                \stackrel{\leftrightarrow}{\partial_0} 
                         B \,d^{n-1} {\mbf x},
  \nonumber 
  \\
&& Q_{\mbox{\scriptsize B(Y)}}
         = -i \int \bar K 
                \stackrel{\leftrightarrow}{\partial_0} 
                         Y_* \,d^{n-1} {\mbf x},
  \nonumber 
  \\
&& Q_{\mbox{\scriptsize B(HK)}}^\prime
         = -i \int \bar K 
                \stackrel{\leftrightarrow}{\partial_0} 
                         B \,d^{n-1} {\mbf x},
  \nonumber 
  \\
&& Q_{\mbox{\scriptsize B(Y)}}^\prime
         = -i \int \bar c 
                \stackrel{\leftrightarrow}{\partial_0} 
                         Y_* \,d^{n-1} {\mbf x},
\label{Eq4:QBs}
\end{eqnarray}
all of which satisfy the nilpotency condition. 
Our BRST charge 
$Q_{\mbox{\scriptsize B}}$ 
can be decomposed as
\begin{equation} 
Q_{\mbox{\scriptsize B}}
   = Q_{\mbox{\scriptsize B(HK)}}
     + Q_{\mbox{\scriptsize B(Y)}}.  
\end{equation} 
The charge $Q_{\mbox{\scriptsize B(HK)}}$  
generates the BRST transformation 
only for the usual fields $\psi_\mu$, $B$, $c$ and $c_*$, 
while 
$Q_{\mbox{\scriptsize B(Y)}}$  
applies only for $Y$, $Y_*$, $K$ and $K_*$.
The charge $Q_{\mbox{\scriptsize B(HK)}}^\prime$  
generates the BRST transformation for $\psi_\mu$ and $B$ but 
with $K$ and $K_*$ treated as their FP ghosts. 
Similarly, 
$Q_{\mbox{\scriptsize B(Y)}}^\prime$  
generates the BRST transformation for $Y$ and $Y_*$
with $c$ and $c_*$  as their FP ghosts. 

In the last section, we have taken (\ref{Eq3:subsidiary}) as 
a physical condition. Instead of it, however, we may 
choose the condition as
\begin{eqnarray}
&& Q_{\mbox{\scriptsize B(HK)}} \left| \mbox{phys} \right\rangle =0,
\nonumber
\\
&& Q_{\mbox{\scriptsize B(Y)}} \left| \mbox{phys} \right\rangle =0.
\label{Eq4:subsidiary2}
\end{eqnarray}
The unphysical modes of gravitino are removed by the first 
equation, while the gaugeon modes by the second. 
We express the space of states satisfying 
(\ref{Eq4:subsidiary2}) by 
${\cal V}_{\mbox{\scriptsize phys}}^{(\alpha)}$. 
As easily seen, this space is a subspace of 
${\cal V}_{\mbox{\scriptsize phys}}$ defined in the last section:
\begin{equation}
    {\cal V}_{\mbox{\scriptsize phys}}^{(\alpha)}
      \subset
       {\cal V}_{\mbox{\scriptsize phys}} .
\end{equation}
We have attached the index $(\alpha)$ to 
${\cal V}_{\mbox{\scriptsize phys}}^{(\alpha)}$
to emphasize that its definition depends on 
the gauge fixing parameter $\alpha$. In fact, 
the BRST charges 
$ Q_{\mbox{\scriptsize B(HK)}} $ 
and
$ Q_{\mbox{\scriptsize B(Y)}}$ 
transform as 
\begin{eqnarray}
&& \widehat Q_{\mbox{\scriptsize B(HK)}}
      = Q_{\mbox{\scriptsize B(HK)}}
        + \tau  Q_{\mbox{\scriptsize B(HK)}}^\prime,
                 \nonumber 
                 \\  
&& \widehat Q_{\mbox{\scriptsize B(Y)}}
   = Q_{\mbox{\scriptsize B(Y)}}
      - \tau  Q_{\mbox{\scriptsize B(HK)}}^\prime,  
\label{Eq4:hatQBs}
\end{eqnarray}
while their sum 
$ Q_{\mbox{\scriptsize B}}$ 
(and thus ${\cal V}_{\mbox{\scriptsize phys}}$) remains invariant. 

Let us define a subspace 
${\cal V}^{(\alpha)}$ 
of the total Fock space ${\cal V}$ by
\begin{equation}
      {\cal V}^{(\alpha)} 
      = \ker Q_{\mbox{\scriptsize B(Y)}} 
   = \{ 
        \left| \Phi \right\rangle 
        \in {\cal V} ; \ 
        Q_{\mbox{\scriptsize B(Y)}} 
                \left| \Phi \right\rangle =0
     \}
      \subset {\cal V}, 
\end{equation}
which includes 
${\cal V}_{\mbox{\scriptsize phys}}^{(\alpha)}$ 
as a subspace since by definition 
${\cal V}_{\mbox{\scriptsize phys}}^{(\alpha)}$ 
can be expressed as 
\begin{equation}
{\cal V}_{\mbox{\scriptsize phys}}^{(\alpha)}
   = \{ 
        \left| \Phi \right\rangle 
        \in {\cal V}^{(\alpha)} ; \ 
        Q_{\mbox{\scriptsize B(HK)}} 
                \left| \Phi \right\rangle =0
     \}
   \subset
      {\cal V}^{(\alpha)} .
\label{Eq4:Vphys(alpha)}
\end{equation}
The space ${\cal V}^{(\alpha)}$ corresponds to the total 
Fock space of the standard formalism in $a=\varepsilon \alpha^2$ gauge. 
And thus, as seen from (\ref{Eq4:Vphys(alpha)}), 
${\cal V}_{\mbox{\scriptsize phys}}^{(\alpha)}$ 
corresponds to the physical subspace 
${\cal V}_{\mbox{\scriptsize phys}}^{\mbox{\scriptsize (HK)}}$ 
of the standard formalism in $a=\varepsilon \alpha^2$ gauge. 
This can be understood from the facts that
\begin{enumerate}
\item The modes of gaugeons and their FP ghosts are removed from 
      the space ${\cal V}^{(\alpha)}$ by the condition 
      $
      Q_{\mbox{\scriptsize B(Y)}} 
      \left| \mbox{phys} \right\rangle =0
      $. 
\item The usual fields $\psi_\mu$, $B$, $c$ and $c_*$ satisfy the 
      (anti)commutation relations exactly the same with those 
      of the standard formalism in $a=\varepsilon \alpha^2$ gauge. 
\end{enumerate}
One may understand the first fact by expressing the 
Lagrangian (\ref{Eq:L_ggon}) as
\begin{eqnarray}
   {\cal L}
       &=&
          {\cal L}_{\mbox{\scriptsize HK}}
                                    {(a=\varepsilon \alpha^2)}
      -i
      \left\{
             Q_{\mbox{\scriptsize B(Y)}}
             , \,
              \partial_\mu \bar K_* \partial^\mu Y
             + i \varepsilon  \bar K_* \dslash
                   \left( {1\over 2}Y_* + \alpha B \right)
      \right\}
        \nonumber \\
      &&
      +\mbox{total derivatives},
\end{eqnarray}
where ${\cal L}_{\mbox{\scriptsize HK}}{(a=\varepsilon \alpha^2)}$ 
denotes the Lagrangian of the $a=\varepsilon \alpha^2$ standard
formalism. 

We emphasize that 
the above arguments are also valid  
if we start from 
the $q$-number gauge transformed charges
(\ref{Eq4:hatQBs}) rather than  
$Q_{\mbox{\scriptsize B(HK)}}$ and 
$Q_{\mbox{\scriptsize B(Y)}}$. 
For example, we can define the subspaces 
${\cal V}^{(\alpha+\tau)}$ and 
${\cal V}_{\mbox{\scriptsize phys}}^{(\alpha+\tau)}$ by 
\begin{eqnarray}
&&  {\cal V}^{(\alpha+\tau)} 
       = \ker \widehat Q_{\mbox{\scriptsize B(Y)}}, 
       \nonumber 
       \\ 
&&  {\cal V}_{\mbox{\scriptsize phys}}^{(\alpha+\tau)}
       = 
         \ker \widehat Q_{\mbox{\scriptsize B(HK)}}
         \cap 
         \ker \widehat Q_{\mbox{\scriptsize B(Y)}}.
\end{eqnarray}
${\cal V}^{(\alpha+\tau)}$ can be identified with the Fock space 
of the standard formalism in $a=\varepsilon (\alpha+\tau)^2$ gauge, 
and 
${\cal V}_{\mbox{\scriptsize phys}}^{(\alpha+\tau)}$ corresponds to
its physical subspace. 
Thus various Fock spaces of the standard formalism 
in different gauges are 
embedded in the single Fock space ${\cal V}$ of 
our theory.\footnote{%
                      Strictly speaking, we have two theories corresponding 
                      to the value of $\varepsilon = \pm 1$. Consequently, 
                      we have two Fock spaces, to which we refer as 
                      ${\cal V}_+$ and ${\cal V}_-$ 
                      corresponding to the value  of $\varepsilon$.  
                      Thus the statement becomes as 
                      follows:  
                      All of the Fock spaces of the standard formalism for 
                      all values of $a \geq 0$ [$a \leq 0$] are 
                      embedded in the single Fock space ${\cal V}_+$ 
                       [${\cal V}_-$] of our theory. 
                    } 

%
%
\section{Comments and discussion}
%
\subsection{Type II theory}
We have seen in \S3 that the gauge fixing parameter 
$\alpha$ can be shifted freely by the $q$-number gauge transformation. 
However, we cannot change the sign of the standard gauge parameter 
$a=\varepsilon \alpha^2$. 
The situation is analogous to the Type I gaugeon 
formalism for QED. There are two types of the gaugeon 
theory for QED.\cite{YK} \ One of them is the Type I theory 
where the standard gauge parameter is expressed as 
$a=\varepsilon \alpha^2$, and the other is the Type II theory 
where the $a=\alpha$. In both types of the theory, $\alpha$ can be 
shifted as $\hat \alpha = \alpha + \tau$ by the $q$-number 
gauge transformation. Thus, in the Type II theory, we can shift 
the standard gauge parameter quite freely. 
We comment here that the Type II theory can be also formulated 
for the  spin-3/2 gauge field. 

Let us consider the following Lagrangian,
\begin{eqnarray}
{\cal L}_{\mbox{\scriptsize II}} 
        &=& {\cal L}_{\mbox{\scriptsize RS}}
          + \bar B \dslash (\gamma \psi) 
          - {i\alpha \over 2} \bar B \dslash B
          - {i \over 2} \bar Y_* \dslash B 
          - \partial_\mu \bar Y_* \partial^\mu Y
          \nonumber \\ &&
          -i \partial_\mu \bar c_*  \partial^\mu c
          -i \partial_\mu \bar K_*  \partial^\mu K.
\label{Eq:L_II}
\end{eqnarray}
Under the $q$-number gauge transformation (\ref{Eq3:qnumbergauge}), 
this Lagrangian is also form invariant (up to total derivatives): 
\begin{equation}
      {\cal L}_{\mbox{\scriptsize II}}(\varphi_A, \alpha)
      =
      {\cal L}_{\mbox{\scriptsize II}}(\hat \varphi_A, \hat \alpha)
      + \mbox{total derivatives},
\label{Eq5:forminvarianceII}
\end{equation}
with $\hat \alpha$ defined by (\ref{Eq3:alphahat}). 
As easily seen, the Lagrangian (\ref{Eq:L_II}) is also invariant (up to 
total derivatives) under all of the transformations corresponding 
to the BRST charges (\ref{Eq4:QBs}). Using the charge 
$Q_{\mbox{\scriptsize B(Y)}}$ we can express the Lagrangian as
\begin{equation}
   {\cal L}_{\mbox{\scriptsize II}}
   ={\cal L}_{\mbox{\scriptsize HK}}{(a=\alpha)}
      -i
      \left\{
             Q_{\mbox{\scriptsize B(Y)}}
             \, , \,
              \partial_\mu \bar K_* \partial^\mu Y
             + 
             {i \over 2}  \bar K_* \dslash B
      \right\}
      +\mbox{total derivatives},
\end{equation}
which leads to the identification 
\begin{equation}
       a=\alpha,
\end{equation}
nothing but the characteristic of a Type II theory. 
It should be noted that 
all the arguments given in \S4 
also apply to this Type II theory.

\subsection{Extended Type I theory}
If we put $\alpha=0$ in the Type I Lagrangian (\ref{Eq:L_ggon}), 
the gaugeon sector decouples from the rest. Then the remaining sector
has the same form with the Lagrangian of the standard formalism in 
the Landau gauge. Thus, the equivalence of the theory to the standard 
formalism is manifest in the Landau gauge. 
This situation does not occur in the Type II theory. 
The gaugeon sector in (\ref{Eq:L_II}) does not decouple for any 
value of $\alpha$. In this sense, the Type I theory is preferable 
to the Type II theory. 
As seen above, however, 
we cannot change the sign of the standard gauge parameter $a$
in the Type I theory, while in the Type II theory we can shift it 
quite freely . 

For the QED case, an extended Type I theory is known,\cite{RE} \  
where we can shift the standard gauge parameter quite freely. 
In this theory, two sets of gaugeons and their FP ghosts are 
introduced. In the following,  we provide an 
extended type I gaugeon formalism for the spin-3/2 gauge field. 

We start from the  Lagrangian,
\begin{eqnarray}
{\cal L}_{\mbox{\scriptsize I}}^\prime 
        &=& {\cal L}_{\mbox{\scriptsize RS}}
          + \bar B \dslash (\gamma \psi) 
          - i (\bar Y_{1*} + \alpha_1 \bar B )
               \dslash 
                ( Y_{2*} + \alpha_2 B )
          - \partial_\mu \bar Y_{1*} \partial^\mu Y_1
          \nonumber \\ &&
          - \partial_\mu \bar Y_{2*} \partial^\mu Y_2
          %
          %
          -i \partial_\mu \bar c_*  \partial^\mu c
          -i \partial_\mu \bar K_{1*}  \partial^\mu K_1
          -i \partial_\mu \bar K_{2*}  \partial^\mu K_2, 
\label{Eq:L_I'}
\end{eqnarray}
where we have introduced two sets of gaugeon fields ($Y_i$,  $Y_{i*}$) and 
their FP ghosts ($K_i$, $K_{i*}$),  
and two gauge fixing 
parameters $\alpha_i$ ($i=1,\,2$). 
If we put $\alpha_1=\alpha_2=0$, then the gaugeon sector decouples 
form the rest and the remaining sector is equal to the Landau 
gauge Lagrangian of the standard formalism. In this sense, this is 
an extension of the Type I theory. 

The Lagrangian (\ref{Eq:L_I'}) is invariant 
up to total derivatives 
under the following BRST transformation:
\begin{eqnarray}
&&   \delta_{\mbox{\scriptsize B}} \psi_\mu
                 = i \partial_\mu c, 
                 \nonumber 
                 \\
&&   \delta_{\mbox{\scriptsize B}} c_* = B, 
                 \nonumber
                 \\
&&   \delta_{\mbox{\scriptsize B}} B
     = \delta_{\mbox{\scriptsize B}} c =0,
                 \nonumber 
                 \\
&&   \delta_{\mbox{\scriptsize B}} Y_i = iK_i, 
                 \nonumber
                 \\
&&   \delta_{\mbox{\scriptsize B}} K_{i*} = Y_{i*}, 
                 \nonumber
                 \\
&&   \delta_{\mbox{\scriptsize B}} Y_{i*} 
      =  
      \delta_{\mbox{\scriptsize B}} K_i = 0, 
                         \qquad (i=1,\,2)
\label{Eq5:I'BRST}
\end{eqnarray}
which satisfies the nilpotency, 
${\delta_{\mbox{\scriptsize B}}}^2=0$. 
The corresponding BRST charge is a sum of 
three nilpotent BRST-like charges:
\begin{equation}
  Q_{\mbox{\scriptsize B}}
  =
  Q_{\mbox{\scriptsize B(HK)}}
  +
  Q_{\mbox{\scriptsize B(Y1)}}
  +
  Q_{\mbox{\scriptsize B(Y2)}},
\label{Eq5:QB}
\end{equation}
where 
$Q_{\mbox{\scriptsize B(HK)}}$ is defined by (\ref{Eq2:QB}) and
$ Q_{\mbox{\scriptsize B(Yi)}}$'s are given by
\begin{equation}
  Q_{\mbox{\scriptsize B(Y$i$)}}
         = -i \int \bar K_i 
                \stackrel{\leftrightarrow}{\partial_0} 
                         Y_{i*} \,d^{n-1} {\mbf x}.
    \qquad (i=1,\,2)
\end{equation}
As usual, the physical subspace is defined by this BRST charge: 
${\cal V}_{\mbox{\scriptsize phys}}=\ker Q_{\mbox{\scriptsize B}}$.

We define the $q$-number gauge transformation by
\begin{eqnarray}
&&   \hat \psi_\mu = \psi_\mu + \tau_1 \partial_\mu Y_1
                              + \tau_2 \partial_\mu Y_2,
                \nonumber \\
&&   \hat Y_{i*} = Y_{i*} - \tau_i B,
                \nonumber \\
&&   \hat B = B, \qquad   \hat Y_i = Y_i,
                \nonumber \\
&&   \hat c = c + \tau_1 K_1 + \tau_2 K_2, 
                \nonumber \\
&&   \hat K_{i*} = K_{i*} - \tau_i c_*,
                \nonumber \\
&&   \hat c_* = c_*, \qquad   \hat K_i = K_i, 
               \qquad (i=1,\,2)
\label{Eq5:I'qnumbergauge}
\end{eqnarray}
where $\tau_i$ 
is the parameter of the transformation
($i=1,\,2$). 
Under this transformation, 
the Lagrangian is form invariant (up to total derivatives): 
\begin{equation}
      {\cal L}_{\mbox{\scriptsize I}}^\prime 
                     (\varphi_A, \alpha_1, \alpha_2)
      =
      {\cal L}_{\mbox{\scriptsize I}}^\prime 
                     (\hat \varphi_A, \hat \alpha_1, \hat \alpha_2)
      + \mbox{total derivatives},
\label{Eq5:I'forminvariance}
\end{equation}
where $\varphi_A$ stands for any of the relevant fields and 
$\hat \alpha_i$'s are defined by 
\begin{equation} 
                 \hat \alpha_i 
                      = \alpha_i + \tau_i. 
                              \qquad (i=1,\,2)
\end{equation} 
The BRST charge
$ Q_{\mbox{\scriptsize B}}$ (\ref{Eq5:QB}) is invariant 
under the $q$-number transformation (\ref{Eq5:I'qnumbergauge}). 
As a result, the physical subspace 
${\cal V}_{\mbox{\scriptsize phys}}$ is gauge invariant. 

To see the relation  of the theory 
to the standard formalism, we may express the  
Lagrangian (\ref{Eq:L_I'}) as 
\begin{eqnarray}
 {\cal L}^\prime_{\mbox{\scriptsize I}}
   &=&
     {\cal L}_{\mbox{\scriptsize HK}}{(a=\alpha_1\alpha_2)}
      -i
      \big\{
             Q_{\mbox{\scriptsize B(Y)}}
             \, , \,
              \partial_\mu \bar K_{1*} \partial^\mu Y_1
              +
              \partial_\mu \bar K_{2*} \partial^\mu Y_2
          \nonumber \\ &&
             + 
             i \bar K_{1*} \dslash Y_2
             +
             i  \alpha_2 \bar K_{1*} \dslash B
             +
             i  \alpha_1 \bar K_{2*} \dslash B
      \big\}
      +\mbox{total derivatives},
\end{eqnarray}
where 
$  Q_{\mbox{\scriptsize B(Y)}}$ is a nilpotent BRST charge 
defined by 
$
  Q_{\mbox{\scriptsize B(Y)}}
  =
  Q_{\mbox{\scriptsize B(Y1)}}
  +
  Q_{\mbox{\scriptsize B(Y2)}}
$. 
This lead us to 
\begin{equation}
                   a=2\alpha_1 \alpha_2, 
\end{equation}
which can be shifted into an arbitrary value 
by the $q$-number gauge transformation 
(\ref{Eq5:I'qnumbergauge}). 

We can show that both of the Fock spaces of 
Type I and Type II theory 
are embedded into the total Fock space of this theory. 
The arguments are quite parallel to the case of QED.\cite{RE} \ 
For example, 
by the $q$-number gauge transformation 
we can always shift the parameters $\alpha_2$ 
into $\alpha_2=1/2$. 
With this value of the parameter, the Lagrangian (\ref{Eq:L_I'}) can 
be expressed as 
\begin{eqnarray}
 {\cal L}^\prime_{\mbox{\scriptsize I}}
   &=&
     {\cal L}_{\mbox{\scriptsize RS}}
      + \bar B \dslash (\gamma \psi)
      -{i \alpha_1 \over 2} \bar B \dslash B 
      -{i \over 2} \bar Y_{1*} \dslash B 
      - \partial_\mu \bar Y_{1*} \partial^\mu Y_1 
      \nonumber \\ &&
      -i \partial_\mu  \bar c_* \partial^\mu c 
      -i \partial_\mu  \bar K_{*1} \partial^\mu K_1 
      \nonumber \\ &&
      -i
      \big\{
             Q_{\mbox{\scriptsize B(Y2)}}
             \, , \,
              \partial_\mu \bar K_{2*} \partial^\mu Y_2
              +
             i (\bar Y_{1*} + \alpha_1 \bar B) \dslash K_{2*}
      \big\},
\end{eqnarray}
which is the same expression of the Type II Lagrangian (\ref{Eq:L_II}) 
up to 
$Q_{\mbox{\scriptsize B(Y2)}}$-exact 
operators. 
Consequently, the subspace 
${\cal V}_{\mbox{\scriptsize II}} 
= \ker 
Q_{\mbox{\scriptsize B(Y2)}}$ can be identified with the 
Fock space of the Type II theory.

\subsection{Gauge invariance}
We have seen in \S4 that 
the subspace 
$
{\cal V}^{(\alpha)}
= \ker
Q_{\mbox{\scriptsize B(Y)}}
\subset
{\cal V}
$ 
can be identified with the total Fock space 
$
{\cal V}^{\mbox{\scriptsize (HK)}}
$
of the standard formalism in $a=\varepsilon \alpha^2$ gauge. 
This does not mean, however, that these spaces are 
isomorphic to each other. Instead, we can show the following 
isomorphism:
\begin{equation}
           {\cal V}^{(\alpha)}
           /
           \,\mbox{Im} \, Q_{\mbox{\scriptsize B(Y)}}
           \cong
           {\cal V}^{\mbox{\scriptsize (HK)}}. 
\label{Eq5:isomorphism}
\end{equation} 
Namely, 
by considering the quotient space
we can ignore the 
the $ Q_{\mbox{\scriptsize B(Y)}}$-exact states (states 
having the form 
$
Q_{\mbox{\scriptsize B(Y)}} | * \rangle
$), 
which have no corresponding states in  
${\cal V}^{\mbox{\scriptsize (HK)}}$. 
Eq.(\ref{Eq5:isomorphism}) is the precise statement that 
our theory includes the standard 
formalism as a sub-theory. 
As for the Hilbert spaces, it can be shown that
\begin{equation}
           {\cal H}^{(\alpha)}_{\mbox{\scriptsize phys}}
           \cong
           {\cal H}^{\mbox{\scriptsize (HK)}}_{\mbox{\scriptsize phys}},
\label{Eq5:isomorphism2}
\end{equation} 
where 
${\cal H}^{(\alpha)}_{\mbox{\scriptsize phys}}$ 
is a physical Hilbert space defined by 
\begin{equation}
   {\cal H}^{(\alpha)}_{\mbox{\scriptsize phys}}
    =
     {\cal V}^{(\alpha)}_{\mbox{\scriptsize phys}}
     /\,
     {\cal N}^{(\alpha)},
\end{equation}
with  
${\cal N}^{(\alpha)}$ being a zero-normed subspace of 
${\cal V}^{(\alpha)}_{\mbox{\scriptsize phys}}$. 
Furthermore, we can also verify that our gauge invariant 
Hilbert space 
${\cal H}_{\mbox{\scriptsize phys}}$
is isomorphic to 
${\cal H}^{(\alpha)}_{\mbox{\scriptsize phys}}$. 
Therefore, we are lead to the gauge invariant result:
\begin{equation}
           {\cal H}^{\mbox{\scriptsize (HK)}}_{\mbox{\scriptsize phys}}
           \cong
           {\cal H}^{(\alpha)}_{\mbox{\scriptsize phys}}
           \cong
           {\cal H}_{\mbox{\scriptsize phys}}.   
\label{Eq5:isomorphism3}
\end{equation} 
The detailed arguments of 
(\ref{Eq5:isomorphism}) and (\ref{Eq5:isomorphism2}) 
will be reported elsewhere. 
Similar discussion  holds for the theories of 
Type II and extended Type I. 

%
\subsection{Background gravitational field}
We have considered so far the theory in the flat space-time. 
It is straightforward to incorporate the interaction 
with the background gravity, if it satisfies Ricci flatness. 

In the background gravitational field $g_{\mu\nu}$, the 
classical Lagrangian (\ref{Eq:L_RS}) becomes
\begin{equation}
{\cal L}_{\mbox{\scriptsize RS}}
                =
           -{i\over 2}\,\sqrt{g} \bar \psi_\mu \gamma^{\mu \nu \lambda}
              D_\nu \psi_\lambda ,
\label{Eq:L_RSg}
\end{equation}
where $g=|\det g_{\mu\nu}|$, $D_\nu$ is the covariant derivative, and 
the Greek indices of $\gamma^{\mu\nu\lambda}$ are now 
of the world coordinate thus having vielbein dependence. 
The Lagrangian (\ref{Eq:L_RSg}) is invariant up to total derivatives 
under the gauge transformation 
\begin{equation}
      \delta \psi_\mu = D_\mu \Lambda, 
\end{equation}
if the background gravitational field satisfy the vacuum Einstein equation: 
\begin{equation}
                        R_{\mu\nu}=0.
\end{equation}

The quantum Lagrangian (\ref{Eq:L_ggon}) is  now given by 
\begin{eqnarray}
{\cal L} &=& {\cal L}_{\mbox{\scriptsize RS}}
          + \sqrt{g} \bar B \Dslash (\gamma \psi) 
          - {i\varepsilon \over 2}\sqrt{g}
                ( \bar Y_* + \alpha \bar B)
                     \Dslash (Y_* + \alpha B)     
          - \sqrt{g}\bar Y_* 
                    \stackrel{\leftarrow}{\Dslash} 
                    \Dslash Y
          \nonumber \\ &&
          -i \sqrt{g} \bar c_*  
                    \stackrel{\leftarrow}{\Dslash} 
                            \Dslash c
          -i \sqrt{g} \bar K_* 
                    \stackrel{\leftarrow}{\Dslash} 
                       \Dslash K.
\label{Eq:L_ggong}
\end{eqnarray}
This Lagrangian is invariant up to total derivatives 
under the BRST transformation 
(\ref{Eq3:BRST}) with an exception for $\psi_\mu$, which
transforms now as 
\begin{equation}
                \delta_{\mbox{\scriptsize B}}
                \psi_\mu
                = i D_\mu c. 
\end{equation}
The corresponding BRST charge is given by 
\begin{equation}
   Q_{\mbox{\scriptsize B}}
   =
   \int \sqrt{g}\, J_{\mbox{\scriptsize B}}^0\,d^{n-1}{\mbf x}, 
\end{equation}
where 
$    J^\mu_{\mbox{\scriptsize B}}$
is the BRST current defined by 
\begin{equation}
    J^\mu_{\mbox{\scriptsize B}}
    = \bar B \stackrel{\leftrightarrow}{D^\mu} c 
                   +
     \bar Y_* \stackrel{\leftrightarrow}{D^\mu} K. 
\end{equation}
(Note that this current is actually conserved since
the fields $\varphi=B$, $Y_*$, $c$ and $K$ satisfy the 
Klein-Gordon equation
$$
 \Box \varphi = D^\mu D_\mu \varphi = 0.
$$
This is due to  the Ricci flatness:
$$
   \Dslash^2 \varphi= \left(D^\mu D_\mu + {1\over 4}R\right)\varphi
        = D^\mu D_\mu \varphi, 
$$
where  $R$ is the scalar curvature $R=g^{\mu\nu}R_{\mu\nu}\ (=0)$.) 
Thus we can consistently define the physical subspace
by 
$
{\cal V}_{\mbox{\scriptsize phys}}=\ker 
Q_{\mbox{\scriptsize B}}
$. 

The form invariance (\ref{Eq3:forminvariance}) also holds 
for the Lagrangian (\ref{Eq:L_ggong}) under the $q$-number 
gauge transformation 
(\ref{Eq3:qnumbergauge})with an exception for $\psi_\mu$, 
which transforms now as
\begin{equation}
  \hat \psi_\mu = \psi_\mu + \tau D_\mu Y.
\end{equation}

All the arguments in \S3 hold also in the present case. 
Especially, the BRST transformation 
commutes with the $q$-number gauge transformation, 
which leads to the gauge invariance of the 
physical subspace (and the physical Hilbert space).

\section*{Acknowledgements}
This work is supported in part by the Grant-in-Aid for 
Scientific Research (C) from the Ministry of 
Education, Science, Sports and Culture (\#08640344).

%


\begin{thebibliography}{99}
%
 \bibitem{N}N.~Nakanishi, Prog.\ Theor.\ Phys.\ Suppl.\ No.51 (1972).
 \bibitem{KO}T.~Kugo and I.~Ojima, Phys.\ Lett.\ {\bf 73B} (1978), 459;
  Prog. Theor. Phys. {\bf 60} (1978), 1896; Prog.\ Theor.\ Phys.\ Suppl.\ 
  No.66 (1979), 1.
 %
 \bibitem{OEM}K.~Yokoyama, Prog.\ Theor.\ Phys.\ {\bf 51} (1974), 1956.
 \bibitem{YK}K.~Yokoyama and R.~Kubo, Prog.\ Theor.\ Phys.\ {\bf 52} (1974), 290.
 \bibitem{txt}K.~Yokoyama, {\it Quantum Electrodynamics} (in Japanese)
                           (Iwanami Shoten, Tokyo, 1978)
 \bibitem{YM1}K.~Yokoyama, Prog.\ Theor.\ Phys.\ {\bf 59} (1978), 1699.\\ 
              K.~Yokoyama, M.~Takeda and M.~Monda, Prog.\ Theor.\ Phys.\ 
                                                  {\bf 60} (1978), 927.\\
              K.~Yokoyama, Prog.\ Theor.\ Phys.\ {\bf 60} (1978), 1167.
 \bibitem{YM2} K.~Yokoyama, M.~Takeda and M.~Monda, Prog.\ Theor.\ Phys.\ 
                                                 {\bf 64} (1980), 1412.
 %
 \bibitem{Smatrix}K.~Yokoyama, Phys.\ Lett.\ {\bf 79B} (1978), 79.
 %
 \bibitem{Abe}M.~Abe, {\it The Symmetries of the Gauge-Covariant Canonical 
               Formalism of Non-Abelian Gauge Theories} 
               (Master Thesis, Kyoto University, 1985). 
 \bibitem{Izawa}K.~Izawa, Prog.\ Theor.\ Phys.\ {\bf 88} (1992), 759.
 \bibitem{KSE}M.~Koseki, M.~Sato and R.~Endo, Prog.\ Theor.\ Phys. \
                     {\bf 90} (1993), 1111.
 \bibitem{KSE_YM}M.~Koseki, M.~Sato and R.~Endo, 
         Bull.\ of Yamagata Univ.,\ Nat.\ Sci.\ {\bf 14} (1996), 15.
 %
 \bibitem{RE}R.~Endo, Prog.\ Theor.\ Phys.\ {\bf 90} (1993), 1121.
%
 \bibitem{HK}H.~Hata and T.~Kugo, Nucl.\ Phys.\ {\bf B158} (1979), 357.
 \bibitem{NK}N.~K.~Nielsen, Nucl.\ Phys.\ {\bf B140} (1978), 499. \\
             R.~E.~Kallosh, Nucl.\ Phys.\ {\bf B141} (1978), 141.
 \bibitem{EK}R.~Endo and T.~Kimura, Prog.\ Theor.\ Phys.\ {\bf 73} (1980), 683.
 %
 \bibitem{ET}R.~Endo and M.~Takao, Prog.\ Theor.\ Phys.\ {\bf73} (1985), 803; 
            Phys.\ Lett.\ {\bf 161B} (1985), 155.

\end{thebibliography}
\end{document}